\documentstyle[aps,prl,epsfig,multicol]{revtex}
\def\bra#1{\left<\,#1\,\right|}
\def\ket#1{\left|\,#1\,\right>}

\begin{document}
\draft

\title{Beam-Induced Nuclear Depolarisation in a 
  Gaseous Polarised Hydrogen Target}

\author{
\centerline {\it The HERMES Collaboration}
K.~Ackerstaff$^{5}$, 
A.~Airapetian$^{32}$, 
N.~Akopov$^{32}$, 
M.~Amarian$^{32,27}$, 
E.C.~Aschenauer$^{13,14}$, 
H.~Avakian$^{10}$, 
R.~Avakian$^{32}$, 
A.~Avetissian$^{32}$, 
B.~Bains$^{15}$, 
C.~Baumgarten$^{23}$,
M.~Beckmann$^{12}$, 
St.~Belostotski$^{26}$, 
J.E.~Belz$^{28,29}$,
Th.~Benisch$^8$, 
S.~Bernreuther$^8$, 
N.~Bianchi$^{10}$, 
J.~Blouw$^{25}$, 
H.~B\"ottcher$^6$, 
A.~Borissov$^{6,14}$, 
J.~Brack$^4$, 
S.~Brauksiepe$^{12}$,
B.~Braun$^{8}$, 
B.~Bray$^3$, 
St.~Brons$^6$,
W.~Br\"uckner$^{14}$, 
A.~Br\"ull$^{14}$, 
E.E.W.~Bruins$^{20}$,
H.J.~Bulten$^{18,25,31}$, 
R.V.~Cadman$^{15}$,
G.P.~Capitani$^{10}$, 
P.~Carter$^3$,
P.~Chumney$^{24}$,
E.~Cisbani$^{27}$, 
G.R.~Court$^{17}$, 
P.~F.~Dalpiaz$^9$, 
E.~De Sanctis$^{10}$, 
D.~De Schepper$^{20}$, 
E.~Devitsin$^{22}$, 
P.K.A.~de Witt Huberts$^{25}$, 
P.~Di Nezza$^{10}$,
M.~D\"uren$^8$, 
A.~Dvoredsky$^3$, 
G.~Elbakian$^{32}$, 
A.~Fantoni$^{10}$, 
A.~Fechtchenko$^7$,
M.~Ferstl$^8$, 
D.~Fick$^{19}$,
K.~Fiedler$^8$, 
B.W.~Filippone$^3$, 
H.~Fischer$^{12}$, 
B.~Fox$^4$,
J.~Franz$^{12}$, 
S.~Frabetti$^9$,
S.~Frullani$^{27}$, 
M.-A.~Funk$^5$, 
N.D.~Gagunashvili$^7$, 
H.~Gao$^{2,15}$,
Y.~G\"arber$^6$, 
F.~Garibaldi$^{27}$, 
G.~Gavrilov$^{26}$, 
P.~Geiger$^{14}$, 
V.~Gharibyan$^{32}$,
V.~Giordjian$^{10}$, 
A.~Golendukhin$^{19,32}$, 
G.~Graw$^{23}$, 
O.~Grebeniouk$^{26}$, 
P.W.~Green$^{1,29}$, 
L.G.~Greeniaus$^{1,29}$, 
C.~Grosshauser$^8$, 
A.~Gute$^8$, 
W.~Haeberli$^{18}$, 
J.-O.~Hansen$^2$,
D.~Hasch$^6$, 
O.~H\"ausser\cite{author_note1}$^{28,29}$, 
R.~Henderson$^{29}$, 
Th.~Henkes$^{25}$,
M.~Henoch$^{8}$, 
R.~Hertenberger$^{23}$, 
Y.~Holler$^5$, 
R.J.~Holt$^{15}$, 
W.~Hoprich$^{14}$,
H.~Ihssen$^{5,25}$, 
M.~Iodice$^{27}$, 
A.~Izotov$^{26}$
% ...
}

\address{
% ...
$^5$DESY, Deutsches Elektronen Synchrotron, 22603 Hamburg, Germany\\
% ...
$^8$Physikalisches Institut, Universit\"at Erlangen-N\"urnberg, 
91058 Erlangen, Germany\\
% ...
$^{14}$Max-Planck-Institut f\"ur Kernphysik, 69029 Heidelberg, Germany\\ 
$^{15}$Department of Physics, University of Illinois, Urbana, 
Illinois 61801, USA\\
% ...
$^{23}$Sektion Physik, Universit\"at M\"unchen, 85748 Garching, Germany\\
% ...
$^{25}$Nationaal Instituut voor Kernfysica en Hoge-Energiefysica (NIKHEF), 
1009 DB Amsterdam, The Netherlands\\
$^{26}$Petersburg Nuclear Physics Institute, St. Petersburg, 188350 Russia\\
$^{27}$Istituto Nazionale di Fisica Nucleare, Sezione Sanita, 
00161 Roma, Italy\\
$^{28}$Department of Physics, Simon Fraser University, Burnaby, 
British Columbia V5A 1S6, Canada\\ 
$^{29}$TRIUMF, Vancouver, British Columbia V6T 2A3, Canada\\
% ...
}
\date{\today}
\maketitle

\begin{abstract}
Spin-polarised atomic hydrogen is used as a gaseous polarised proton
target in high energy and nuclear physics experiments operating with
internal beams in storage rings.  When such beams are intense and bunched,
this type of target can be depolarised by a resonant interaction
with the transient magnetic field 
generated by the beam bunches. This effect has been studied  with the 
HERA positron beam in the HERMES experiment at DESY. Resonances have 
been observed and  a simple analytic model has been used to explain 
their shape and position. Operating conditions for the experiment have 
been found where there is no significant target depolarisation due to 
this effect.
\end{abstract}

\begin{multicols}{2}[]

Nuclear-polarised hydrogen and deuterium gas targets deployed in high energy 
storage rings have become an important tool in the study of spin dependent 
processes in nuclear and particle physics experiments. They offer a 
unique combination of large nucleon polarization with the absence of 
other polarized or unpolarised nuclear species. A potentially serious 
practical consideration in the use of this type of polarised target in 
bunched beams is the nucleon depolarisation which can take place when 
the transient magnetic fields generated by the beam interact with the 
polarised nucleons and change their spin state. This can occur when 
the frequency $f_h$ of any large amplitude harmonic in the frequency spectrum 
of the pulsed magnetic field produced by the beam bunch structure coincides 
with an atomic hyperfine transition frequency $\nu$ determined by the local 
value of the static magnetic field which provides the quantisation axis for 
the nucleon spins. To minimise this depolarisation it is therefore necessary 
to provide static magnetic field conditions which ensure that no such resonant
effects can occur within the effective target volume. These resonant
depolarisation processes can be studied experimentally only with a fully
operational target installed in a suitable storage ring.  Beam induced
depolarisation effects have been previously observed with tensor polarised 
deuterium targets \cite{gilman,ferro-luzzi} in low energy electron beams in 
the VEPP-3 and NIKHEF accelerators. In this Letter we report on the first 
observation and measurements of beam-induced resonant depolarisation using 
a hydrogen target with a high energy positron beam in the HERMES experiment 
at DESY.

The HERMES experiment at DESY uses deep inelastic scattering to study the spin 
structure of the nucleon with polarised internal targets and a 27.5~GeV 
high intensity polarised positron beam in the HERA storage ring.
The proton target is generated by injecting a nuclear-polarised atomic 
hydrogen beam from an atomic beam source (ABS) into a tubular open-ended 
storage cell which confines the atoms in the region of the circulating beam. 
The storage cell increases the probability of a positron-proton interaction 
by a factor of approximately one hundred compared with the free atomic beam. 
This results in a target thickness that is useful for HERMES with the atomic 
beam fluxes ($\sim 7\times10^{16}$  atoms s$^{-1}$) which are currently
available from polarised atomic hydrogen beam sources. 
A static magnetic field of approximately 335 mT directed parallel to the 
positron beam axis is provided throughout the target cell to define a 
longitudinal quantisation axis for the proton spins and to decouple the 
atomic electron and proton spins.

The HERA positron beam is optimised for collider operation and therefore 
consists of very short bunches with high peak currents.  The beam has a time
structure which allows for up to 220 bunches, which have a length of 27~ps 
(1$\sigma$) and are separated by a time interval of 96~ns corresponding 
to a bunch frequency $f_b$~=~10.4~MHz. The beam cross section at the target 
region is elliptical with a height of 0.07~mm and a width of 0.31~mm 
(1$\sigma$). Typically, the beam current has a maximum value of 45 mA 
immediately after injection, and is allowed to decay to 15 mA before the 
beam is dumped and the storage ring is refilled \cite{HERA}.
The frequency spectrum of the 
transient field has been calculated with the assumption that the magnetic 
field of each bunch has an approximately Gaussian shape in time. Hence the 
spectrum has the form of a harmonic series with a Gaussian shaped amplitude envelope. 
The bunch length is very short compared with the bunch separation; 
therefore the width of this envelope is large (6 GHz at 1$\sigma$) 
and very high-numbered harmonics have significant amplitudes.

\begin{figure}
 \begin{center}
 \epsfig{file=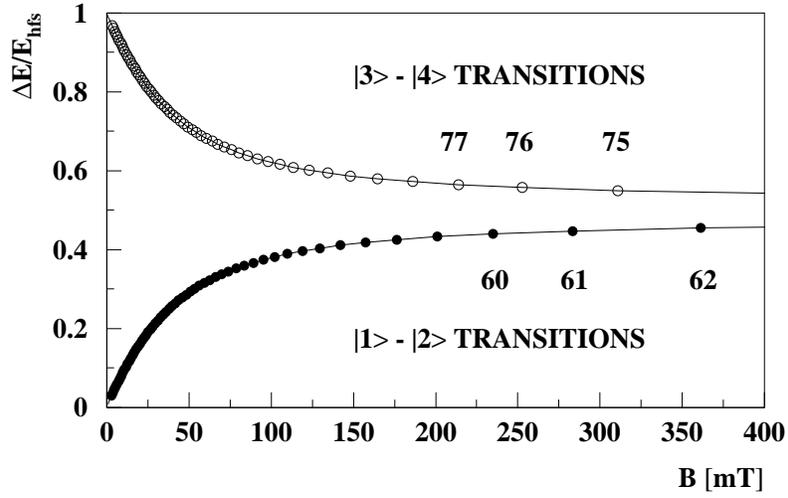, width=12cm}
 \end{center}
  \caption{The calculated energy difference in units of $E_{HFS}$ (1420 MHz)
    for the $\ket{1}$ -- $\ket{2}$
    and $\ket{3}$ -- $\ket{4}$ transitions as a
    function of static magnetic field. Each point shows
    the position of a proton depolarising resonance in the HERA beam.
    The harmonic numbers ($k=f_h/f_b$)
    for the relevant resonances are given.}
  \label{deltaE}
\end{figure}

In a magnetic field the hydrogen atom has four substates 
which are $\ket{m_I,m_J}=$ $\ket{+\frac{1}{2},+\frac{1}{2}}$,
$\ket{-\frac{1}{2},+\frac{1}{2}}$,
$\ket{-\frac{1}{2},-\frac{1}{2}}$ and
$\ket{+\frac{1}{2},-\frac{1}{2}}$, labelled $\ket{1}$ to $\ket{4}$ respectively.
In normal target operation states $\ket{1}$ and $\ket{4}$ are selected to 
give positive (spin parallel to the static field) and $\ket{2}$ and $\ket{3}$ 
to give negative (spin anti-parallel) proton 
polarisations. In both cases the total
electron polarisation is small.  Loss of proton 
polarisation can occur if the resonance 
condition for either transition $\ket{1}$ -- $\ket{2}$ or 
$\ket{3}$ -- $\ket{4}$ is satisfied.    
The positions of the proton depolarising 
resonances as a function of static magnetic field can 
be deduced from the magnetic sub-state energies and have been
calculated for the HERA beam bunch frequency
as shown in Fig.~\ref{deltaE}.

To avoid beam-induced depolarisation in normal target operation, the value
and the spatial uniformity of the static field must be chosen so that
there are no resonances within the effective target volume \cite{kinney}. The resonance
separation increases as the field value is raised and it follows that the 
fractional uniformity requirement can be most easily satisfied if the
highest possible value for the static field is used.   
However, practical considerations impose a limit 
on the maximum available field of approximately 
350~mT for the HERMES target when operating in 
longitudinal mode. The spacing of adjacent 
resonances is approximately 50~mT in this region. 
Therefore, the target field magnet was  
designed to have a maximum value of 400~mT with 
a uniformity better than $\pm2\%$ over the target 
cell volume.

\begin{figure}
 \begin{center}
 \epsfig{file=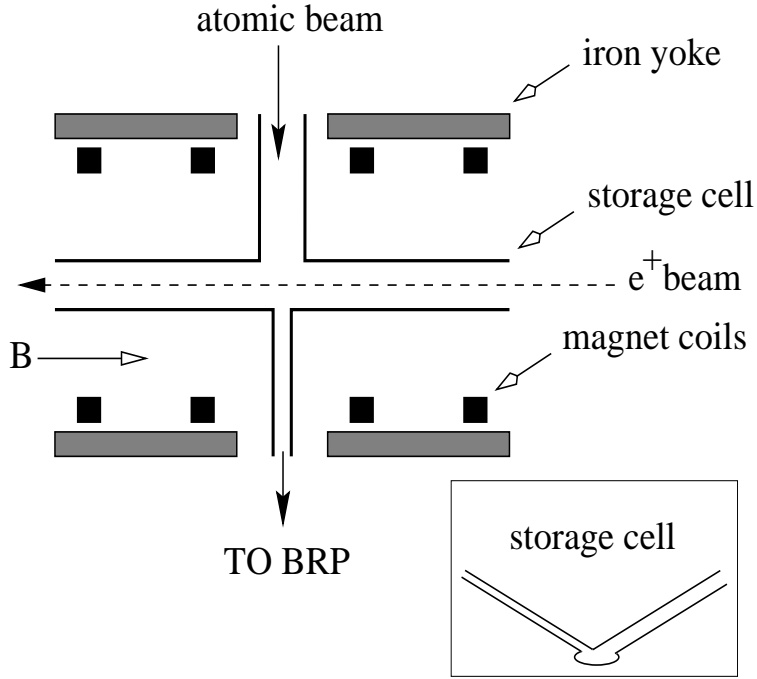, width=10cm, height=9cm}
 \end{center}
  \caption{A schematic layout of the HERMES hydrogen target. 
    The inset shows the cross section of the storage cell 
    at its centre point.}
  \label{target}
\end{figure}

\begin{figure}
 \begin{center}
 \epsfig{file=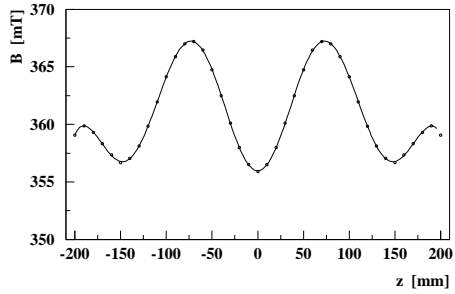, width=7cm}
 \end{center}
  \caption{The static magnetic field profile plotted as a function
    of axial position ($z$).
    The points are measurements to which the line is fitted.
    The maxima at $z = \pm 75$ mm and $\pm 190$ mm  correspond
    approximately to the $z$ positions of the coils.}
  \label{fprof}
\end{figure}

The basic layout of the target is shown in Fig.~\ref{target}. 
The nuclear-polarised atomic beam from the ABS \cite{gol}
is injected via a side tube into the centre ($z=0$)
of the target cell, which is 400~mm long and 
has an elliptical cross section 29~mm wide by 
9.8~mm high \cite{stewart}. The atoms diffuse to the open 
ends of the cell at thermal velocities, generating a triangular density 
distribution along the beam axis. The escaping gas is removed from the
storage ring beam tube by high speed vacuum pumps. The cell is cooled
to approximately 100~K to maximise the target thickness, which is 
$7\times10^{13}$ atoms cm$^{-2}$ under normal operating conditions. 
The static field is provided by four superconducting coils mounted coaxially
to the HERA positron beam, with a measured field profile along the positron
beam axis as shown in Fig.~\ref{fprof}.

The polarisation of a sample of the atoms in the cell is measured 
with a Breit-Rabi polarimeter (BRP) \cite{braun}. The sample 
flows out of a second side tube also connected to the centre of the cell.
The BRP uses adiabatic RF transition units in combination with 
a sextupole magnet system, and a quadrupole mass 
spectrometer (QMS) incorporating an atom-counting detector
to measure the magnetic sub-state populations. 
The nuclear and electronic polarisations may then 
be calculated from the population values obtained.

\begin{figure}
 \begin{center}
 \epsfig{file=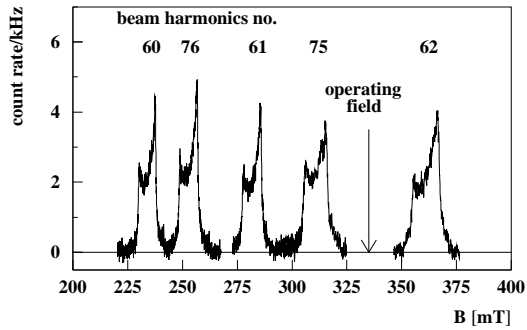, width=8cm}
 \end{center}
  \caption{A composite plot of the observed resonances as a function
    of static magnetic field in the range 220 to 400~mT shown
    with the corresponding beam harmonic number. The vertical
    axis is the count rate for hydrogen atoms detected by the 
    QMS.}
\label{fscan}
\end{figure}

The resonances were observed experimentally by making measurements with
the BRP over a range of static magnetic field values    
in the region of the normal operating field (335~mT), while the stored 
HERA positron beam was circulating through the target storage cell. Two
different observation techniques were used. In the first approach, 
the ABS operated in the normal way and atoms in both states $\ket{1}$ 
and $\ket{4}$ were injected into the target cell. 
The BRP was then arranged so that states $\ket{1}$ and $\ket{4}$ 
were not detected by the QMS, with the result that any signal 
seen in this detector was due to the presence of 
states $\ket{2}$ or $\ket{3}$. A statistically significant signal 
could be obtained in a short time with this technique, 
so it was possible to scan the field continuously and observe directly the 
position and shape of any resonant depolarisation signal. 
Measurements were made with stored positron beam currents between 38 and 28~mA.
The field was scanned continuously over a range
covering the full width of each resonance. A composite
plot of the QMS signals produced by each of the resonances that occur in
the field range between 220 and 400~mT is shown in Fig.~\ref{fscan}. This 
shows that states other than $\ket{1}$ and $\ket{4}$ were being populated 
in the cell, clearly demonstrating the presence of resonant depolarisation 
effects. The observed resonances are at the expected values 
of the magnetic field within the experimental uncertainties.

\begin{figure}
 \begin{center}
 \epsfig{file=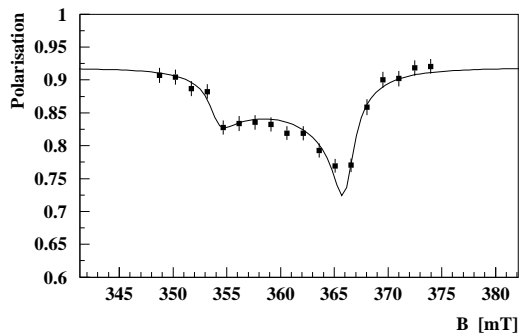, width=8cm}
 \end{center}
  \caption{The measured proton polarisation as a function
    of static magnetic field in the region 
    of the 62$^{\mathrm nd}$ beam harmonic.
    The fitted line is described in the text.}
  \label{sres}
\end{figure}

The second method involves measuring the polarisation of the sampled atoms in 
the normal way by  determining the individual populations of each of the
four possible states in the sampled atomic beam. The proton polarisation 
could then be deduced from these data. This technique required the static
magnetic field to be constant while the state populations for 
each point were measured. The resonance produced by the  $k = 62$
harmonic was studied in detail and the results are plotted in Fig.~\ref{sres},
which demonstrates the loss in the polarisation produced 
by this particular resonance.

The data were compared with a simple analytic model incorporating certain
basic features of atomic kinetics in the target cell as well as
the expected dependencies of the atomic transition probability on the time
structure of the beam transient field as modulated by the atom's passage 
through the beam \cite{kolster}. Hence the shape of the resonances depends 
on the atomic density distribution within the cell, the static 
magnetic field profile and the BRP sampling sensitivity. 
A number of approximations were used to greatly simplify this many-dimensional
problem. The fundamental
approximation is that the effects of the beam fields on the atoms can be
calculated with an {\em average} transition probability, 
which is appreciable only in close proximity to the beam, 
and at locations where the static magnetic field magnitude $B(z)$ 
is at a resonant value. If the depolarisation is small so that each 
depolarised atom can be uniquely associated
with the location in $z$ where it was depolarised, one can assume 
that the rate of depolarised atoms $N_D$ observed by the BRP can be calculated
from the flux of atoms which enter the beam interaction region at position $z$,
$\phi(z)$, weighted by both the average transition probability $W_{ab}(B(z))$ 
at that location and the probability $S(z)$ that any flux at position $z$ 
will be sampled at the BRP:
\[N_D= \int_{cell} W_{ab}(z) \phi(z) S(z) \,{\rm d}z.\]
Then ${P_m=P_0(1-N_D/N)}$, where $P_m$ is the measured polarisation, $P_0$
is the polarisation in the absence of the HERA beam and $N$ is the total rate
of atoms entering the BRP. The average transition probability $W_{ab}$ is
calculated to first order assuming that the perturbing transient field is 
purely transverse relative to the static field, and neglecting the effect 
of the changing vector direction of the transient field as the atom passes 
through the beam \cite{abragam}. The transition probability
\[W_{ab}(z)=\int_0^\infty |\,\bra{b}\hat\mu\cdot\vec 
B_t \ket{a}\,|^2\, \rho(\nu(z) -f)\, {\cal F}^2(f) \,{\rm d}f.\]
The frequency distribution $\rho$ is assumed to be a 
Lorentzian of width $\Gamma$ that corresponds to an average time
$\tau$ of passage of an atom through the effective beam interaction
region. ${\cal F}(f)$ is the calculated Fourier transform function and $f$ the
frequency of the transient field. Only one harmonic in $\cal F$ is relevant 
for any given resonance, and its line shape depends on the distribution of the
bunch intensities in the stored beam. The operator $\hat\mu$ contains the
Pauli matrices for the electronic and nuclear spins which operate on the 
eigenstates $\ket{a}$ and $\ket{b}$ of the atom in the static field,  
with strength given by the appropriate magnetic dipole moment.
In general, the positron beam will produce a transient field with 
amplitude $B_{t}$ which has a value increasing with the radial distance 
$r$ inside the beam and falling externally as $1/r$. 
To simplify the analysis, $B_t$ is approximated by a
constant throughout a hypothetical cylindrical interaction volume
situated on the beam axis, and zero outside. The width $\Gamma$ varies
inversely with the diameter of this volume.

In practice the prediction of the model is normalised to the data by an
arbitrary factor which incorporates the perturbative field strength $B_t$ 
and the ratio $\phi(0)/N$.
The width $\Gamma$ is a second fit parameter. A third fit parameter 
accounts for the relative contribution from the direct atomic beam compared 
to the diffusive flow of atoms.
The flux $\phi(z)$ has a triangular shape because the
atoms are fed in at the centre of the cell.
The sampling probability $S(z)$ decreases linearly with the distance from
the sampling tube and therefore has an identical shape. 
Fig.~\ref{sres} illustrates how well the model can be fitted to the 
measured polarisation data. The best fit value of $\Gamma$ gives 
$\tau$~=~0.98~$\mu$s, which corresponds to an interaction path 
length of 2.5~mm when the atoms are travelling  
with a thermal velocity  corresponding to a temperature 
of 100~K. This value is compatible with the size 
of the effective interaction region.

The double peak structure can be simply understood by considering
the profile in $z$ of the static field (Fig.~\ref{fprof}), as resonance 
depolarisation is enhanced when it occurs where the gradient of the static 
field is small. When the magnet excitation current is high enough to place 
the central minimum in the field profile at the resonance value, a 
depolarisation peak is produced. The proximity of this resonance region to the
sampling tube, together with the enhancement from direct flow, results 
in a larger depolarisation signal than in the case of the peak at lower 
magnet excitation, which arises when the resonant field value occurs in the
two field maxima at $|z|=$~75~mm. The model result for the relative magnitude
of these two peaks is sensitive to the third fit parameter.

In summary, we have observed depolarising resonances in the HERMES hydrogen 
target as expected from initial design studies made for the 
experiment. 
The features of the observed resonances are accounted for 
by a simple analytic model based on 
the target cell geometry, the profile of the target 
magnetic field and the nominal beam parameters.
A working point for the target static field has 
been found between two resonances, where the 
beam-induced target depolarisation is undetectable at an  
uncertainty level below 1\%. This demonstrates 
that a gaseous polarised proton target can be 
operated without significant beam-induced 
depolarisation effects in a stored positron beam 
having the very high bunch density required for collider operation.

We gratefully acknowledge the DESY management for its support,
the DESY staff and the staffs of the collaborating institutions 
for their significant effort and our funding agencies for financial support.

%
%

\end{multicols}
%
%
\end{document}